\documentclass[twocolumn, superscriptaddress]{revtex4-2}

\usepackage[utf8]{inputenc}
\usepackage[T1]{fontenc}
\usepackage[english]{babel}
\usepackage[dvipsnames]{xcolor}
\usepackage{lmodern,fancyhdr}
\usepackage{graphicx,float}
\usepackage{enumitem,amsmath,amssymb,amsthm,bm,array,mathdots,mathrsfs,dsfont,bbold}
\usepackage[colorlinks]{hyperref}
\hypersetup{colorlinks,linkcolor=blue,citecolor=blue,urlcolor=blue,final}
\DeclareMathAlphabet\mathbfcal{OMS}{cmsy}{b}{n}
\usepackage{comment}
\usepackage{siunitx}
\usepackage{booktabs}
\newcommand{\ra}[1]{\renewcommand{\arraystretch}{#1}}

\begin{document}

\hyphenpenalty=5000
\tolerance=1000

%\title{(tbd) Optically levitated spinning top: Rotation of a nanudumbbell around its long axis}
\title{Spinning a levitated mechanical oscillator far into the deep-strong coupling regime}
% \title{Optically levitated spinning top} Lukas' suggestion

\author{J.~A.~Zielińska}
\email{jzielinska@eth.ch}
\affiliation{Photonics Laboratory, ETH Zürich, CH-8093 Zürich, Switzerland}

\author{F.~van der Laan}
\affiliation{Photonics Laboratory, ETH Zürich, CH-8093 Zürich, Switzerland}
\affiliation{Center for Nanophotonics, AMOLF, 1098 XG Amsterdam, The Netherlands}

\author{A.~Norrman}
\affiliation{Photonics Laboratory, ETH Zürich, CH-8093 Zürich, Switzerland}
\affiliation{Center for Photonics Sciences, University of Eastern Finland, P.O. Box 111, FI-80101 Joensuu, Finland}

\author{R. Reimann}
\affiliation{Photonics Laboratory, ETH Zürich, CH-8093 Zürich, Switzerland}
\affiliation{Quantum Research Center, Technology Innovation Institute, Abu Dhabi, UAE}

\author{M. Frimmer}
\affiliation{Photonics Laboratory, ETH Zürich, CH-8093 Zürich, Switzerland}

\author{L.~Novotny}
\affiliation{Photonics Laboratory, ETH Zürich, CH-8093 Zürich, Switzerland}

% \author[1]{J.~A.~Zielińska}
% \email{jzielinska@eth.ch}
% \author[1,2]{F.~van der Laan}
% \author[3]{A.~Norrman}
% \author[1,4]{R.~Reimann}
% \author[1]{M.~Frimmer}
% \author[1]{L.~Novotny}

% \affil[1]{Photonics Laboratory, ETH Zürich, CH-8093 Zürich, Switzerland}
% \affil[2]{Center for Nanophotonics, AMOLF, 1098 XG Amsterdam, The Netherlands}
% \affil[3]{Institute of Photonics, University of Eastern Finland, P.O. Box 111, FI-80101 Joensuu, Finland}
% \affil[4]{Quantum Research Center, Technology Innovation Institute, Abu Dhabi, UAE}

\begin{abstract}

The field of levitodynamics has made substantial advancements in manipulating the translational and rotational degrees of freedom of levitated nanoparticles. Notably, rotational degrees of freedom can now be cooled to millikelvin temperatures and driven into GHz rotational speeds. However, in the case of cylindrically symmetric nanorotors, only the rotations around their short axes have been effectively manipulated, while the possibility to control rotation around the longer axis has remained a notable gap in the field.
Here, we extend the rotational control toolbox by engineering an optically levitated nanodumbbell in vacuum into controlled spinning around its long axis with spinning rates exceeding 1 GHz. This fast spinning introduces deep-strong coupling between the nanodumbell's libration modes, such that the coupling rate $g$ exceeds the bare libration frequencies $\Omega_0$ by two orders of magnitude with $g/\Omega_0=724\pm 33$.
Our control over the long-axis rotation opens the door to study the physics of deep-strong coupled mechanical oscillators and to observe macroscopic rotational quantum interference effects, thus laying a solid foundation for future applications in quantum technologies.
Additionally, we find that our system offers great potential as a nanoscopic gyroscope with competitive sensitivity. 

\end{abstract}

\maketitle

\section*{Introduction.} 

Cylindrically symmetric nanorotors, when trapped within linearly polarized optical tweezers, align their most polarizable (long) axis to the polarization direction of the tweezers.
This results in the emergence of two degenerate libration modes corresponding to oscillatory rotations around the two least-polarizable (short) axes. Recently, significant advances have been made in controlling the short axis rotations, with nanorotors being driven to spin at GHz rates~\cite{ReimannPRL2018, AhnPRL2018, AhnNatNano2020, vanderLaanPRA2020}, libration oscillations cooled down to mK temperatures~\cite{BangPRR2020,vanderLaanPRL2021,SchaeferPRL2021,Pontin2023}, and numerous demonstrations of torque sensing~\cite{AhnNatNano2020, Xu2017PRA, KuhnNatComm2017, RashidPRL2018,ju2023nearfield}. 
However, the remaining rotational degree of freedom, which involves the nanoparticle's rotation around its long axis, typically remains unconstrained and is driven by thermal fluctuations~\cite{SebersonPRA2019, BangPRR2020, vanderLaanPRL2021}. 
Introducing controlled long axis spinning as a tool within rotational levitodynamics would enable the study of deep-strong coupling between mechanical modes, open up an avenue towards engineered quantum rotational dynamics~\cite{SticklerNature2021}, and significantly improve real-world technological prospects of nanoscopic levitated gyroscopes.

The generation of strongly and ultrastrongly coupled  states is an ongoing experimental effort of the levitodynamics community~\cite{Magrini18Optica, DelosRiosSommer2021, Rieser2022Science, dare2023linear}, as it paves the way towards non-Hermitian dynamics~\cite{rudolph2023quantum} and entanglement generation~\cite{VitaliPRL2007}. 
%While cavity-based mechanical systems face limitations in achieving increased couplings due to unstable dynamics~\cite{Peterson2019, dare2023linear}, an intriguing aspect arises when the coupling emerges naturally in rotational mechanics. The coupling between two libration modes of a cylindrically symmetric rotor, caused by long-axis spinning motion, can become arbitrarily large without introducing mechanical instabilities, which is made possible by the dissipative nature of the coupling. 
%The deep-strong coupling regime, in which the coupling rate $g$ dominates over all other system dynamics, including the frequencies $\Omega_0$ of the uncoupled system, is currently most impressively realized in trapped atoms with ${g/\Omega_0=6.5}$~\cite{Koch2023}. 
The deep-strong coupling regime is reached when the coupling rate of two modes $g$ dominates over all other system dynamics, including the eigenfrequencies of the uncoupled system.
%becomes comparable to their eigenfrequencies. 
%deep-strong coupling regime, in which the coupling rate $g$ dominates over all other system dynamics, including the frequencies $\Omega_0$ of the uncoupled system, is currently most impressively realized in trapped atoms with ${g/\Omega_0=6.5}$~\cite{Koch2023}. 
However, reaching such strong coupling is not trivial since it can result in unstable dynamics~\cite{Peterson2019, dare2023linear}. Interestingly, we find that the deep-strong coupling regime can be readily reached for the libration modes of a cylindrically symmetric rotor. Here, the coupling rate is proportional to the long-axis spinning frequency, which can be made arbitrarily large without running into mechanical instabilities. We reach coupling strengths $g$ that are a factor of 724 larger than the uncoupled libration frequencies $\Omega_0$, outperforming the  $g/\Omega_0$ values demonstrated with trapped atoms by a large margin~\cite{Koch2023}.
%The ultrastrong coupling regime is reached as the coupling rate $g$ becomes comparable to the frequencies $\Omega_0$ of the uncoupled sub-components. Further increasing control, one enters the deep-strong coupling regime in which $g$ exceeds $\Omega_0$~\cite{Forn-Diaz2019}. 
%Although our comprehension of the deep-strong coupling regime is still in its infancy, it is believed that gaining control over systems within that regime will yield various applications in modern quantum technologies~\cite{FriskKockum2019, Forn-Diaz2019}. 
\begin{figure}[t]
    \centering
    \includegraphics[width=0.3\textwidth]{"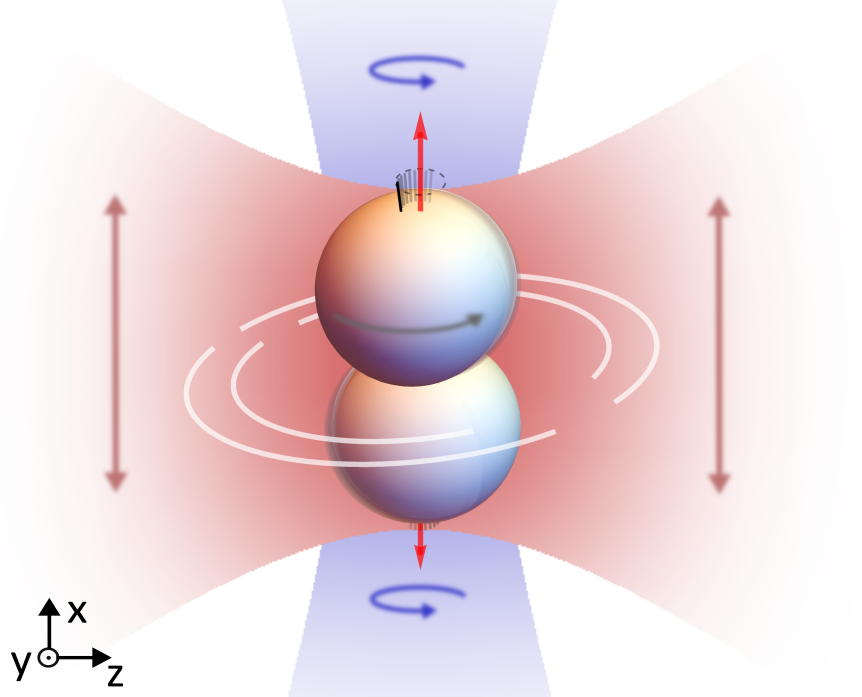"}
    \caption{A nanodumbbell (not to scale) is trapped by an $x$ polarized laser beam (red) and spun around its long axis by a circularly polarized spinning beam (blue).}
    \label{fig:top}
\end{figure}

%Novel experimental techniques that enhance system control often unlock future technologies in quantum- and nanomechanics.
%Control can be quantified by comparing the coupling rate $g$ between specific sub-components of the system, with characteristic system parameters.

%In the strong coupling regime, $g$ exceeds decoherence or loss rates, which provides the needed control for many recent scientific and engineering breakthroughs~\cite{Dovzhenko2018}.  
% In this work, we experimentally realize a controllable rotation of a nanodumbbell around its long axis. The rotation, reaching over a GHz rate, is inferred from the coupling it creates between the two libration degrees of freedom. The coupling strength reaches a value two orders of magnitude larger than the natural libration frequency. 

%In addition, q
One of the main objectives of rotational levitodynamics is to probe quantum interference in the orientational dynamics of nanorotors, which have no analogue in the centre-of-mass motion of the object and no correspondence in classical physics~\cite{SticklerNature2021}. %The fundamental difference between quantum physics of rotors and harmonic oscillators is already evident in the energy level structure and algebraic form of the commutation relations between the angular momentum components. 
%Non-commutative nature of angular momentum components of a free rigid rotor gives rise to uncertainty relation between components of angular momentum $J_1$, $J_2$ and $J_3$, which reads $\Delta J_1 \Delta J_2\ge \frac{1}{2}\hbar |\langle J_3 \rangle|$~\cite{opticallypolarizedatoms}, where $\Delta$ denotes standard deviations. 
The angular momentum of a free quantum rotor, e.g., a molecule, is quantized and in principle it can be prepared in a single eigenstate~\cite{KochRevModPhys2019}. 
%Currently, the mainstream effort in the rotational levitodynamics field is focused on orienting (cooling) a rotor in 3D~\cite{vanderLaanPRL2021, Pontin2023}. This, however, mimics the harmonic oscillator physics (since an aligned state is a superposition of many angular momentum components in a basis of a free rotor eigenstates). 
Unlike librational modes, which are governed by harmonic dynamics, the spinning around the long axis defines a \textit{free rotational} degree of freedom and its corresponding angular momentum component is quantized. Controlling long-axis spinning while simultaneously manipulating (e.g. by feedback cooling) and detecting the remaining two angular directions allows to create a state with a large main angular momentum component.
%Controlled long-axis spinning demonstrated here allows us to create a state with a large main angular momentum component while efficiently manipulating (e.g. feedback cooling) and detecting the remaining two angular directions.
This in turn paves the way towards observing angular momentum quantization on a macroscopic scale and eventually engineering a nanorotor in a well-defined angular momentum quantum state. 
%\textcolor{red}{Controlled long-axis spinning allows us to create a state with a large main angular momentum component while efficiently manipulating and detecting the remaining two angular directions. This capability opens the door to measure shot noise resulting from the quantization of angular momentum. Such a measurement would mark the first experimental evidence of macroscopic quantum rotational mechanics.}
%These may include observation of symmetric top-like energy levels structure~\cite{KochRevModPhys2019} and   
What is more, exploring the interaction between rotations around the long axis and light is a notable experimental and theoretical gap in the field. 
Understanding the resulting decoherence of nanorotors is crucial for the exploration of quantum rotational effects~\cite{SticklerNature2021}, such as rotational revivals~\cite{SticklerNJP2018} or rotational quantum non-demolition measurements~\cite{usami2007quantumnoiselimited}. 
% Typically, theoretical models of light-induced decoherence for symmetric rotors use “stick” (one-dimensional object whose orientation is completely described by two angles) Hamiltonians in the small-angle approximation~\cite{SticklerNJP2018}. 
% These models fail to predict precisely the experimentally measured light-induced reheating rate for librating nanodumbbells~\cite{vanderLaanPRL2021}. This is hardly surprising, since the length-to-diameter ratio for nanodumbbells is approximately 1.8 (they clearly are not 1D objects). On top of the limitations of current models of decoherence for libration, to our knowledge no sources currently offer estimates of the light-induced recoil heating rate suffered by the non-librational degree of freedom (spinning around long axis). 
% %and generation minimum uncertainty coherent angular momentum states~\cite{Atkins1971}, 
% Understanding of recoil heating rate of nanorotors is key for exploration of quantum rotational effects~\cite{SticklerNature2021} such as rotational revivals~\cite{SticklerNJP2018}. Additionally, there's an enticing possibility a scheme analogous to spin quantum non-demolition (QND) measurment may be revealed ~\cite{usami2007quantumnoiselimited}.

%(Additionally this makes for convenient platform for spin-rotational coupling experiments mediated by magnetic field.) 

Finally, fast-spinning levitated nanoparticles also offer significant advantages for inertial sensing of rotations. These objects, isolated from the surrounding environment, can be controlled down to their fundamental quantum and thermodynamic limits~\cite{vanderLaanPRA2020,vanderLaanPRL2021}. 
%In contrast, conventional mechanical gyroscope technologies often face limitations due to technical issues. 
However, for gyroscope applications, it is essential to achieve efficient detection of the spinning axis orientation, which is not feasible when the particle spins around its short axis~\cite{TebbenjohannsPRA2022}. Consequently, to fully explore the advantages of levitated nanoparticles for inertial sensing, it is vital to attain controlled rotation around the long axis.

In this work, we experimentally realize controlled rotation of a nanodumbbell around its long axis. 
The rotation rate, reaching over a GHz, is inferred from the coupling it creates between the two libration degrees of freedom.
This coupling reaches far into the deep-strong coupling regime with ${g/\Omega_0=724\pm 33}$.

%Finally, improving experimental control over strongly coupled degrees of freedom is currently an important effort in the levitodynamics field~\cite{dare2023linear, Rieser2022Science, Magrini18Optica}, as it paves the way towards non-hermitian dynamics~\cite{rudolph2023quantum} and entanglement generation~\cite{VitaliPRL2007}. While experimental platforms such as cavity-based systems face limitations in achieving strong coupling due to unstable dynamics~\cite{dare2023linear}, an intriguing aspect arises when the coupling emerges naturally in rotational mechanics. The coupling between two libration modes of a cylindrically symmetric rotor caused by the long-axis spinning motion can be arbitrarily large without introducing mechanical instabilities. This is made possible by the dissipative nature of the coupling, which enters via velocities rather than positions of the oscillators. Such a feature may prove valuable for frequency separation between libration and center-of-mass (COM) dynamics, diminishing the cross-coupling effects on the COM phase space distribution~\cite{kamba2023observation}.

\section*{System under study.} Our system, illustrated in Fig.~\ref{fig:top}, consists of an optically levitated nanodumbbell exposed to a light field in a three-dimensional (3D) polarization state~\cite{NorrmanJEOS2017,AlonsoAOP2023}. The light field is composed of two parts: a strong tweezers field, linearly polarized along the $x$ axis and propagating in the $z$ direction, and a weak spinning beam, circularly polarized in the $yz$ plane and propagating along the $x$ direction. The tweezers field exerts a conservative restoring torque on the dumbbell, which aligns its long axis in the $x$ direction and creates two libration modes around this equilibrium. In contrast, the circularly polarized beam generates a non-conservative torque, which drives the dumbbell into a spinning motion around its long axis. This torque arises from the optical spin~\cite{BliokhPR2015,GilPRA2023} carried by the beam together with absorption or imperfect cylindrical symmetry of the dumbbell~\cite{BellandoPRL22, FonsThesis, kamba2023nanoscale}.

The rotational dynamics of the nanodumbbell~\cite{supplementary} is thus governed by two libration angles $\varphi$ and $\vartheta$ with respect to the $x$ axis and the rotation angle $\psi$ around its long axis. 
In the presence of friction and assuming small-angle libration, the dumbbell’s spinning rate will accelerate until it reaches its steady-state value ${\dot{\psi}_{0}=\tau/(I_3\gamma_3}$)~\cite{vanderLaanPRA2020, ReimannPRL2018, AhnPRL2018}, where $\tau$ is the magnitude of the non-conservative torque, $I_3$ is the moment of inertia along the long axis, and $\gamma_3$ is the friction coefficient. The spinning motion reveals itself as a coupling $g= (I_3/2 I_1) \dot{\psi}_{0}$ between the otherwise independent and harmonic libration modes, where $I_1$ is the moment of inertia along one short axis.
%Let us now focus on the rotational dynamics of the nanodumbbell and the role this non-conservative torque plays in it. We describe the deviations of the dumbbell's long axis from laboratory $x$ axis using two angles $\varphi$ and $\vartheta$, corresponding to the libration modes of the system~\cite{supplementary}. % in $xy$-plane and $xz$-plane respectively~\cite{supplementary}. The remaining angular degree of freedom corresponding to the rotation of the dumbbell around its long axis is denoted by $\psi$. In the presence of friction (described by coefficient $\gamma_3$), the dumbbell's spinning rate will accelerate until it reaches its steady-state value $\dot{\psi}_{0}=k/(I_3\gamma_3$)~\cite{vanderLaanPRA2020, ReimannPRL2018, AhnPRL2018}, where $I_3$ is the moment of inertia for rotation around the long axis.
%In the small angle approximation, for a non-spinning dumbbell the libration modes $\varphi$ and $\vartheta$ are independent harmonic oscillators. The spinning motion reveals itself as an additional coupling $g= (I_3/I_1) \dot{\psi}_{0}$ between the libration modes $\varphi$ and $\vartheta$.
%according to the following equations of motion~\cite{SebersonPRA2019}:
%\begin{subequations}
%\begin{align}
% \ddot{\vartheta} +\omega_\mathrm{s} \dot{\varphi}=  -  \Omega_0^2 \vartheta \;,\\
%  \ddot{\varphi} -\omega_\mathrm{s} \dot{\vartheta} = - \Omega_0^2 \varphi\;,
%    \label{eq:coupled}
%\end{align}
%\end{subequations}
%where $\omega_\mathrm{s}=(I_3/I_1) \dot{\psi}_{0}$ and $\Omega_0$ is the natural libration frequency. 
The coupling leads to hybrid modes with eigenfrequencies
\begin{equation}
\Omega_{1/2}=\sqrt{ \Omega_0^2+g^2}\pm g ,
\label{eq:precfrs}
\end{equation}
where $\Omega_0$ is the natural libration frequency~\cite{SebersonPRA2019}. The higher frequency $\Omega_{1}$ corresponds to nutation of the long axis, whereas the lower frequency $\Omega_{2}$ is associated with precession of the long axis around the $x$ direction~\cite{Landau1976Mechanics}.
\begin{figure}[t]
    \centering
    \includegraphics[width=0.9\linewidth]{"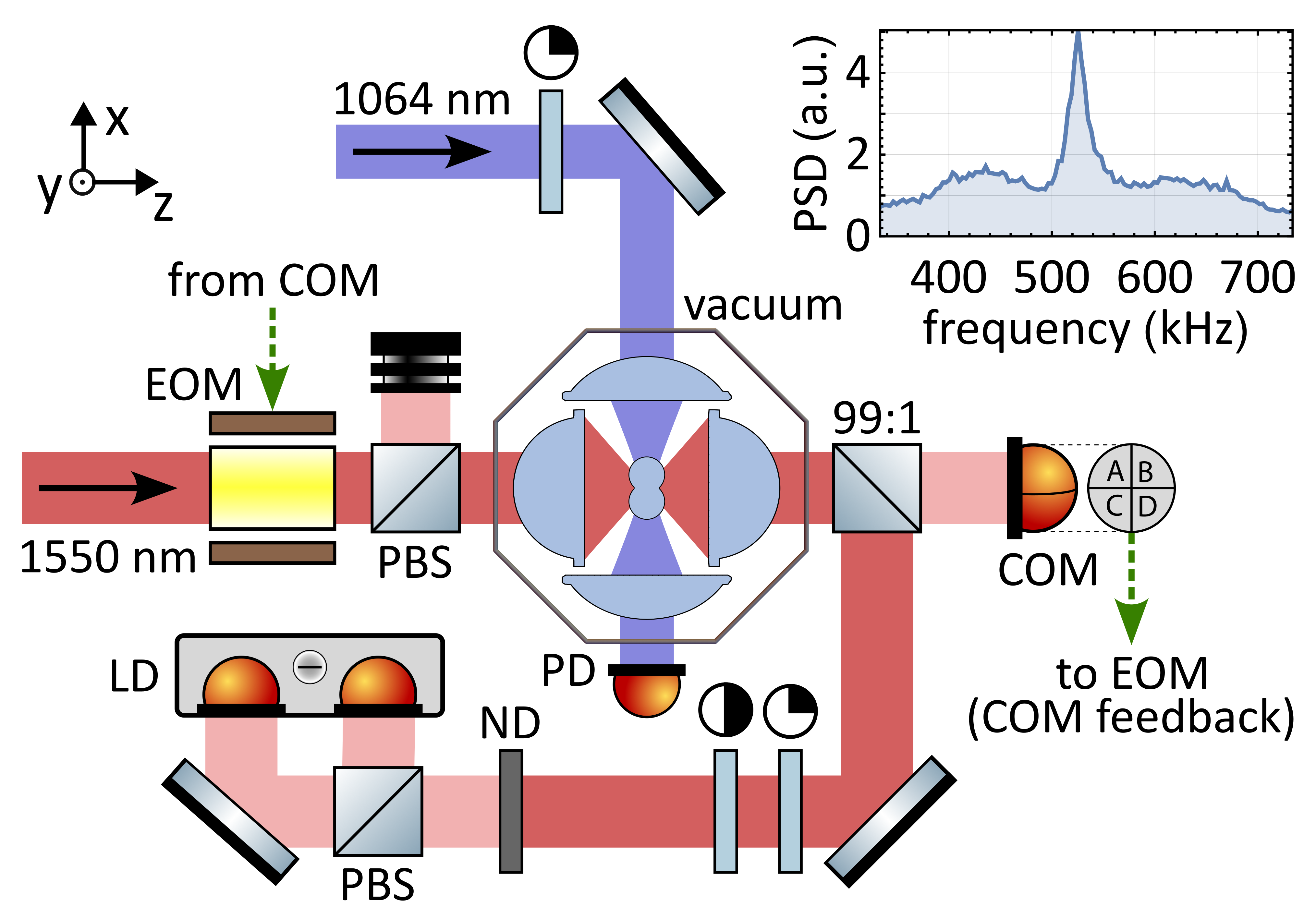"}
    \caption{Experimental apparatus for spinning a dumbbell. The $x$ polarized optical tweezers (wavelength \SI{1550}{\nm}) trap a dumbbell inside a vacuum chamber, while a circularly polarized beam (wavelength \SI{1064}{\nm}) propagating along $x$ spins the dumbbell. Forward scattered light from the tweezers is split and subsequently sent to a quadrant photodiode and a balanced detector for center-of-mass (COM) and libration detection (LD), respectively. The detected COM motion is used to drive an electro-optic modulator (EOM) for feedback cooling  $x$ and $y$ COM motion. The power of the spinning beam is monitored on a photodiode (PD). Inset: power spectral density (PSD, plotted in arbitrary units) of the libration motion of a trapped dumbbell recorded at a pressure of $0.1~{\rm mbar}$ and without the spinning beam.}
    \label{fig:setup}
\end{figure}
%$\Omega_{1}$ and $\Omega_{2}$ correspond to counterclockwise and clockwise precession motion of the long axis around the $x$ axis.
%\begin{figure}[b]
%    \centering
%    \includegraphics[width=0.44\textwidth]{"PrecFreqNatUnits.pdf"}
 %   \caption{Precession modes $\Omega_1$ and $\Omega_2$ calculated from Eq.~\eqref{eq:precfrs} in the units of natural libration frequency $\Omega_0$.}
 %   \label{fig:prectheo}
%\end{figure}
%The dependence of $\Omega_{1/2}$ on the coupling rate $\omega_\mathrm{s}$ is shown in Fig.~\ref{fig:prectheo}.
\begin{figure*}[t]
\includegraphics[width=\textwidth]{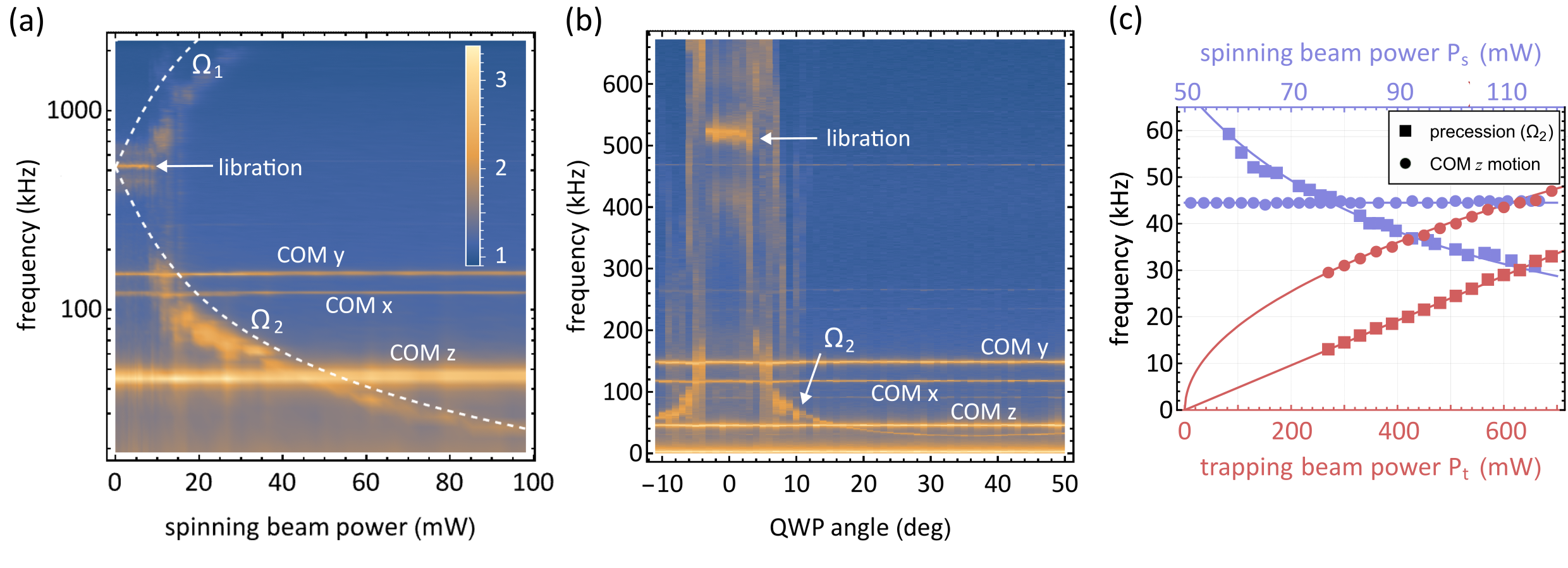}
\caption{Measured effects of spinning on the libration power spectral density (PSD).  Color in plots (a) and (b) shows $\log (\rm{PSD})$ in arbitrary units. Measurements (a)-(c) were performed at a pressure of $10^{-3}~\rm{mbar}$. (a): Measured PSDs as a function of the spinning beam power.
%, illustrating unstable dynamics observed in the slow spinning regime. 
As we increase the spinning beam power the libration peak splits into the high frequency ($\Omega_1$) nutation mode and low frequency ($\Omega_2$) precession mode, which are separated by $2g$. Fits to Eq.~\eqref{eq:precfrs} are shown as white dashed lines. 
%Inadvertent detection of COM motion appears as horizontal lines.  
(b): Measured PSDs for different polarizations of the spinning beam with optical power of \SI{100}{\mW}. 
%At the quarter-wave plate (QWP) angle \SI{0}{\degree} the spinning beam is linearly polarized and has no effect on the libration. Around the QWP angle \SI{45}{\degree} the spinning beam is circularly polarised, and instead of libration spectrum we observe precession mode frequency $\Omega_2$ below COM $z$ frequency. 
As the quarter-wave plate (QWP) angle is adjusted from \SI{0}{\degree} to \SI{45}{\degree}, the spinning beam polarization changes from linear to circular, and instead of the libration spectrum we observe the precession mode with frequency $\Omega_2$ which dips below the COM $z$ frequency. 
(c): Precession frequency $\Omega_2$ and COM $z$ motion frequency as a function of trapping beam power (with a \SI{100}{\mW} circularly polarized spinning beam) and as a function of spinning beam power (with a \SI{700}{\mW} trapping beam). 
%COM $z$ motion frequency is shown with square root and linear fits for trapping and spinning beam power dependence, respectively. In turn, precession frequency is shown together with fits according to Eq.~\eqref{eq:slowmode}.
Fits to Eq.~\eqref{eq:slowmode} are shown as solid lines.
}
\label{fig:SpinVsParams}
\end{figure*}

Most importantly, since the coupling rate $g=\tau/(2 I_{1}\gamma_{3})$ is directly related to the torque and friction, with the torque scaling linearly with the spinning beam power~\cite{supplementary}, in our experimental implementation we can control the coupling via the optical power of the circularly polarized spinning beam and the chamber pressure. 
For high enough spinning rates, we can push our system far into the deep-strong coupling regime, i.e., $g\gg\Omega_0$.
%which is the focus of this work. 
In this regime we can approximate the eigenfrequencies as
%For increasing coupling $g$, the precession frequency $\Omega_2$ tends towards zero, whereas the nutation frequency $\Omega_1$ increases proportional to $g$. Without the constant torque $k$, the coupling is generated by thermal fluctuations of $\dot{\psi}$, leading to $g$ to also fluctuate in time around $0$. However, in our experimental implementation the coupling rate $g$ is not fixed, but we are able to manipulate it by controlling the circularly polarized beam and pressure in the vacuum chamber. For high enough spinning rates, we can bring our system into the deep strong coupling regime, i.e. when $g\gg\Omega_0$, which we will explore in the remainder of this work. In this regime we can approximate the eigenfrequencies as:
\begin{subequations}
\begin{align}
\Omega_{1}&= 2g \;,\label{eq:fastmode}\\
\Omega_2&=\frac{\Omega_0^2}{2g} \label{eq:slowmode}.
\end{align}
\end{subequations}
 %The dumbbell's motion is a combination of a slow precession of the long axis around the $x$ axis with frequency $\Omega_{2}$ and a fast nutation with frequency $\Omega_{1}$ in opposite direction~\cite{Landau1976Mechanics}.
 As we increase the spinning rate of the dumbbell, the amplitude of nutation relative to precession decreases, and the motion becomes dominated by slow precession with frequency $\Omega_2$~\cite{Goldstein}. At the same time, the value of $\Omega_2$ depends only weakly on $g$ for high spinning rates.

\section*{Experimental setup.}

In our experiment, depicted schematically in Fig.~\ref{fig:setup}, silica nanodumbbells composed of two spherical nanoparticles (nominal diameter $143~{\rm nm}$) are optically trapped inside a vacuum chamber. The tweezers beam forming the trap (wavelength $1550~{\rm nm}$, power $700~{\rm mW}$) is focused with an ${\rm NA}=0.8$ lens. The light scattered by the particle is subsequently collected by an ${\rm NA}=0.7$ lens and analyzed to detect libration in the $xy$ plane and the center-of-mass (COM) degrees of freedom. For detecting the COM motion we use a quadrant photodiode, while libration is recorded with the help of a balanced detector as in~\cite{vanderLaanPRL2021}. In order to stabilize the position of the dumbbell inside the trap, we cool the $x$ and $y$ COM motion at pressures below $10^{-4}$~mbar by parametric feedback~\cite{Gieseler2012}. The cooling is realized using an electro-optic modulator (EOM), which modulates the tweezers' power at twice the $x$ and $y$ COM frequencies~\cite{Gieseler2012}.

We identify a trapped particle as a dumbbell when its COM $x$-to-$y$ gas damping ratio is in the $1.1-1.15$ range~\cite{AhnPRL2018}. 
Another characteristic of dumbbells is their libration spectrum, which consists of a sharp libration peak at $\SI{525(3)}{\kHz}$ flanked by broad shoulders, as shown by the inset of Fig.~\ref{fig:setup}. 
This spectral shape originates from coupling between the libration modes introduced by thermally driven spinning around the long axis~\cite{SebersonPRA2019, BangPRR2020, vanderLaanPRL2021, ZielinskaPRL2023}.

The spinning beam (wavelength \SI{1064}{\nm}, tunable power up to \SI{120}{\mW}) is focused onto the dumbbell by an ${\rm NA}=0.3$ lens with $7.5$~mm focal length. The beam is then collected by another ${\rm NA}=0.3$ lens and its power is monitored using a photodiode. The polarization state of the spinning beam is controlled by a quarter-wave plate (QWP). 
%Note that the spinning beam alone would align the dumbbell in the $yz$ plane. However, because of low power and weak focusing, it does not significantly alter the libration potential.
Due to its low power and weak focusing, the spinning beam does not significantly alter the libration potential~\cite{supplementary}.

\begin{figure*}[t]
\includegraphics[width=\textwidth]{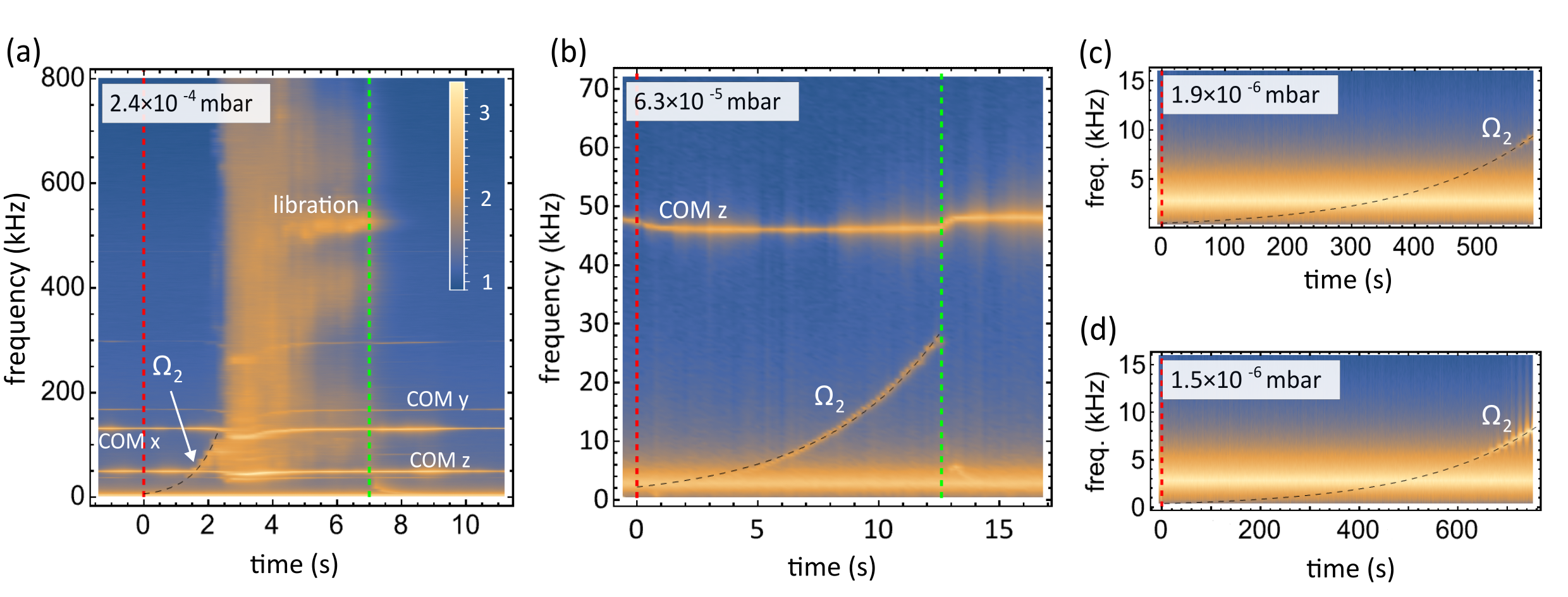}
\caption{Ringdown measurements at different pressures.  Each subfigure shows PSDs measured by the libration detector as a function of time $t$, after blocking the spinning beam (power $50~\rm{mW}$) at $t=0$ marked by a dashed red line. 
The nanodumbbell's spinning rate decreases due to friction, which is evidenced by the increase of the precession mode frequency $\Omega_2$. 
The time the spinning beam is switched back on is indicated in subfigures (a) and (b) by a dashed green line. Color in all plots shows $\log (\rm{PSD})$ in arbitrary units.
Black dashed lines are the fitted precession mode frequency, according to Eq.~\eqref{eq:ringdown}. 
%The scale bar for all subfigures is shown in the lower right corner of the figure. 
{\bf (a):} Measurement at $2.4 \times 10^{-4}~{\rm mbar}$, without parametric feedback cooling of the COM. After approximately $4~{\rm s}$ the dumbbell stops spinning and the libration peak appears. 
{\bf (b)-(d):} Measurement with parametric feedback cooling of the COM $x$ and $y$ motion at $6.3\times 10^{-5}~{\rm mbar}$,  $1.9 \times 10^{-6}~{\rm mbar}$, and $1.5 \times 10^{-6}~{\rm mbar}$, respectively.
%The spinning beam is switched back on after $12.5~{\rm s}$. 
With decreasing pressure, the timescale for slowing down the free but damped spinning motion increases and the precession mode starts to become visible at later times. 
%{\bf (c):} As the dumbbell's spinning slows down, the precession mode only starts to be visible after $400~{\rm s}$ since blocking the spinning beam. 
%{\bf (d):} Ringdown measurement at $1.5 \times 10^{-6}~{\rm mbar}$, with parametric feedback cooling of the COM $x$ and $y$ motion. As the dumbbell's spinning slows down, the precession mode only starts to be visible after $550~{\rm s}$ since blocking the spinning beam. After this particular experiment particle loss occured at around $800~{\rm s}$.
{\bf (d):} After this particular experiment the particle is lost at around $800~{\rm s}$.
}
\label{fig:ringdown}
\end{figure*}
 
\section*{Results.} Figure~\ref{fig:SpinVsParams}(a) shows the measured libration spectra for different optical powers of the circularly polarized spinning beam at $\SI{e-3}{\milli\bar}$. For powers below \SI{10}{\mW} the thermally-driven spinning dominates, as evidenced by the constant libration frequency accompanied by the shoulder-like lineshape on both sides (cf. inset in Fig.~\ref{fig:setup}). 
%As we increase the spinning beam power above $10~{\rm mW}$ threshold, the libration dynamics visibly splits into two hybrid modes created by the spinning motion. 
As we increase the spinning beam power above the $10~{\rm mW}$ threshold, the libration mode splits into two hybrid modes created by the spinning motion. 
%However, the hybrid mode frequencies are unstable until the spinning beam power reaches $40~{\rm mW}$.
The white dashed curves are theoretical fits according to Eq.~\eqref{eq:precfrs} and agree well with the experimental observations. 
%The behavior predicted by Eq.~\eqref{eq:precfrs}, according to which the coupling rate $g$ is proportional to the spinning beam power, is shown as white dashed lines.
We observe a broad high frequency nutation peak ($\Omega_1$) and a narrow low frequency precession mode ($\Omega_2$). 
As expected, the frequency $\Omega_1$ of the nutation mode increases and its amplitude decreases~\cite{Goldstein} with increasing spinning beam power. 
For spinning beam powers below $40~{\rm mW}$ our measurements deviate from the theoretical model, which we ascribe to thermal and pressure-induced fluctuations, coupling to COM motion, and to deviations of the dumbbell  from perfect cylindrical symmetry.
%our model does not accurately predict the frequencies, as well as other features, such as experimentally observed unstable behavior and the $10~{\rm mW}$ spinning beam power threshold for mode splitting. We believe these deviations could be related to the factors not included in our model, such as thermal fluctuations of the spinning rate, pressure fluctuations, coupling to the COM motion, or imperfect cylindrical symmetry of the dumbbell. 
However, we obtain very good agreement for spinning beam powers above $40~{\rm mW}$, where the spinning rate of the dumbbell transitions into the fast spinning regime characterized by ${g\gg\Omega_0}$ [see also Fig.~\ref{fig:SpinVsParams}(c)].

%Additionally, we recognize that the width of the $\Omega_1$ peak (nutation) is broader than that of the $\Omega_2$ peak (precession), which reflects the fact that $\Omega_1$ is more sensitive to thermal fluctuations of the spinning rate than $\Omega_2$ [see Eqs.\eqref{eq:fastmode}--\eqref{eq:slowmode}].

The torque $\tau$ experienced by the dumbbell can also be controlled by the polarization of the spinning beam. In Fig.~\ref{fig:SpinVsParams}(b) we show the measured libration spectra as a function of the QWP angle that changes the polarization of the spinning beam from linear to circular (see Fig.~\ref{fig:setup}). When the QWP is set at $0 ^\circ$, the beam is linearly polarized in the $z$ direction and has no detectable effect on the libration dynamics due to lack of optical spin. The coupling is dominated by thermally driven spinning, as if the spinning beam was absent. However, when the polarization becomes elliptical (QWP is rotated by approx. 5 degrees in either direction) the dynamics transitions abruptly into consistent spinning, as evidenced by the emerging low frequency precession mode ($\Omega_2$). For circular polarization (around $45^\circ$), $\Omega_2$ decreases clearly below the COM $z$ frequency visible at $45~\rm{kHz}$.
%Due to long duration of this measurement, pressure drift might have affected the exact position of the minimum $\Omega_2$ (maximum spinning rate). 

Finally, we investigate the dependence of the precession frequency $\Omega_2$ on the trapping beam power $P_{\rm t}$ and the spinning beam power $P_{\rm s}$ in the fast spinning regime (${g\gg\Omega_0})$. 
The natural libration frequency $\Omega_0$ was shown to depend on the square root of $P_{\rm t}$~\cite{SebersonPRA2019}. This implies, according to Eq.~\eqref{eq:slowmode}, that the precession frequency $\Omega_2$ depends linearly on $P_{\rm t}$, as long as the coupling rate $g$ is not affected by the trapping beam. The expected behavior is indeed confirmed by our measurements shown in Fig.~\ref{fig:SpinVsParams}(c). On the other hand, since the torque $\tau$ exerted on the dumbbell depends linearly on the spinning power $P_{\rm s}$ we expect, according to Eq.~\eqref{eq:slowmode}, that $\Omega_2$ scales inversely with $P_{\rm s}$. We compare the behaviour of $\Omega_2$ 
%as a function of powers $P_{\rm t}$ and $P_{\rm s}$ 
with that of the eigenfrequency of the COM $z$ mode (which is expected to follow a square-root behavior on $P_{\rm t}$ and remain unaffected by changes in the spinning beam power $P_{\rm s}$). The measured $\Omega_2$ and COM $z$ frequencies shown in Fig.~\ref{fig:SpinVsParams}(c) agree well with their respective predicted behaviors. We therefore conclude that Eq.~\eqref{eq:slowmode} correctly predicts the precession frequency in the fast spinning regime where $g\gg\Omega_0$.
%According to Eq.~\eqref{eq:slowmode}, $\Omega_2$ depends linearly on the trapping power since it is proportional to the square of the natural libration frequency. We plot $\Omega_2$ as a function of trapping beam power in Fig.~\ref{fig:SpinVsParams}(c). The fit demonstrates that the data indeed follows the expected linear relation. 
%This result shows that the coupling rate $g$ (and consequently the spinning rate $\dot{\psi}_0$) does not depend on the trapping beam power, which determines the amplitude of libration around the laboratory $x$ axis. 
%We contrast the linear dependence on trapping beam power, with the behaviour of the COM motion. In the same figure we also plot the eigenfrequency of the $z$ COM mode, which follows a square-root behaviour (as indicated by the fit) with trapping power.
Note that the coupling strength surpasses the natural libration frequency even for moderate $P_{\rm s}$. This is evidenced by the data in Fig.~\ref{fig:SpinVsParams}(c), which shows that the precession frequency $\Omega_2$ decreases to \SI{30}{\kHz} for $P_{\rm s}\approx \SI{100}{\mW}$. According to Eq.~\eqref{eq:slowmode}, this corresponds to a coupling rate ${g=2\pi\times \SI{4.6}{\MHz}}$. 
%more than an order of magnitude larger than the natural libration frequency. 
Assuming our dumbbells have a length-to-diameter ratio of 1.8 (as in Ref.~\cite{BellandoPRL22}), we can estimate $I_3/I_1\approx 0.6$, which yields a spinning rate of $\dot{\psi}_0\approx 2\pi\times\SI{15}{\MHz}$.

%In summary, the data in Figs.~\ref{fig:SpinVsParams}(a)-(c) indicate that we are able to drive the spinning far beyond thermally excited motion and, once in the fast spinning regime, we have excellent control over the rate of rotation of the dumbbell around its long axis.
%We set the torque $k$ driving the spinning by adjusting the power and the polarization of the spinning beam, resulting in deep-strong coupling for easily accessible system parameters.

\section*{Ringdown measurements.} 
%Having demonstrated how tuning the optical torque $k$ affects the libration dynamics, we will now employ the damping rate $\gamma_3$ to manipulate the coupling rate. T
The steady-state spinning rate $\dot{\psi}_0=\tau/(I_3\gamma_3)$ is inversely proportional to the damping rate $\gamma_3$, which in turn is proportional to the gas pressure. 
Thus, controlling the pressure allows us to tune $\dot{\psi}_0$ over several orders of magnitude. 
As described before, we can infer the dumbbell's spinning rate from the measured hybrid mode frequencies. However, in practice we are only able to detect the precession mode $\Omega_2$ for spinning rates up to  $\dot\psi_0 \approx \SI{30}{\MHz}$. 
The reason for this is two-fold: first, as $\Omega_2/(2\pi)$ decreases below $3~\rm{kHz}$ it becomes obscured by electronic noise in our detection system;
%second, because of angular momentum conservation large spinning rates reduce the area of the precession peak area leading to poor signal-to-noise in our measurements~\cite{supplementary}.
second, large spinning rates lead to large angular momenta which stabilize the system and reduce the precession amplitude which results in poor signal-to-noise in our measurements~\cite{supplementary}.
%Our detection efficiency limits the peak area we are able to discern and thus sets an upper limit on the spinning rate we can directly infer.
We circumvent this problem by performing ringdown measurements  by setting $P_{\rm s}$ to zero and observing the slowing down of the dumbbell's spinning rate~\cite{ReimannPRL2018}.
%The spinning beam is switched off at time $t=0$, after which the spinning rate  will exponentially decay to zero~\cite{ReimannPRL2018}. 
Once the spinning rate approaches zero
%, changes in the spinning rate will be dominated by 
thermal fluctuations become dominant and the ringdown is a stochastic trajectory. 
However, for sufficiently high spinning rates 
%much larger than typical thermal motion (which results in spinning rate fluctuations on the order of \SI{100}{\kHz}) the trajectory is accurately described by a deterministic ringdown. 
%Therefore, we model the spinning rate as 
thermal fluctuations can be ignored and the ringdown turns into a deterministic trajectory described by ${\dot{\psi}(t)= \dot{\psi} (0) e^{-\gamma_3 t}}$. Consequently, according to Eq.~\eqref{eq:slowmode}, the precession frequency  exponentially increases in time as:
\begin{equation}
\Omega_2(t)= \frac{\Omega_0^2}{g(0)} e^{\gamma_3 t}.
\label{eq:ringdown}
\end{equation}
Note that this description is only valid when $g \gg \Omega_0$, and thus we only use Eq.~\eqref{eq:ringdown} to determine $g(0)$ and $\gamma_3$ in this parameter regime. 
%using data before the coupling rate approaches the natural libration frequency.

Figure~\ref{fig:ringdown}(a) shows a ringdown measurement performed at a pressure of $2.4 \times 10^{-4}~\rm{mbar}$. The spinning beam is switched off at $t=0$ (red dashed line) and the dumbbell's spinning slows down until it reaches the thermally-driven regime at $t=\SI{4}{\s}$ where the libration motion becomes visible. The beam is switched back on after approximately $7~\rm{s}$ (green dashed line) and the random thermal motion gives way to the controlled spinning again. We fit the data with Eq.~\eqref{eq:ringdown} (black dashed line) for times up to $\SI{2.2}{\s}$ and observe that $\Omega_2$ increases exponentially as expected. We present the obtained damping and coupling rate together with the spinning rate in Table~\ref{tab:results}.

At pressures below $10^{-4}~\rm{mbar}$ we parametrically feedback cool the $x$ and $y$ COM modes for stabilizing the particle's position in the trap to avoid particle loss.
% However, this poses certain limitations on the ringdown measurement. First, the pump-down has to be performed in presence of the spinning beam. Second, the spinning beam needs to be switched back on before the precession mode reaches COM frequencies and impacts the PLL (which inevitably causes particle loss).
%As the COM frequencies are perturbed when the precession frequency is similar (see Fig.~\ref{fig:ringdown}(a) at $t=3~{\rm s}$), we switch the spinning beam back on before $\Omega_2$ approaches the COM motion frequencies, as not to disturb the feedback cooling and prevent loosing the particle.
As an example, we show a ringdown measurement at \SI{6.3e-5}{\milli\bar} in Fig.~\ref{fig:ringdown}(b).
The cooling is performed by modulating the trapping beam at twice the COM $x$ and $y$ frequencies~\cite{Gieseler2012}.
The spinning beam is switched back on at $t\approx \SI{12.5}{\s}$, before $\Omega_2$ reaches the COM $z$ frequency. The data again show that $\Omega_2$ increases exponentially after switching off the spinning beam, but the increase is slower than in Fig.~\ref{fig:ringdown}(a) because of the lower pressure. 
Note that $\Omega_2$ is only discernible from \SI{4}{\s} onwards, where it emerges from the low frequency noise. 
%From a fit to Eq.~\eqref{eq:ringdown}, shown again as a black dashed line, we extract the coupling rate $g(0)=2\pi \times \SI{63(2)}{\MHz}$ and damping rate $\gamma_3=\SI{204(2)}{\mHz}$. 
Using this method we can measure high values of $g$, even when $\Omega_2$ is not detectable in steady-state, i.e., when the spinning beam is continuously on.

\begin{table*}\centering
\ra{1.3}
\begin{tabular}{p{0.07\textwidth} p{0.12\textwidth} p{0.17\textwidth} p{0.18\textwidth} p{0.15\textwidth} p{0.09\textwidth}}
    \toprule[1.5pt]
     Fig. 4 & $p_\mathrm{gas}$ (mbar) & $\gamma_3$ (Hz) & $\mathbf{g(0)/(2\pi )\,(MHz)}$ & $\dot{\psi}(0)/(2\pi )$ (MHz) & $g(0) / \Omega_0$\\
    \midrule[1pt]
    (a) & \num{2.4E-4} & \num{7.69(6)E-1} & $\mathbf{2.25(25)}$ & \num{7.5(8)} & \num{4.29(48)}\\
    (b) & \num{6.3E-5} & \num{2.045(29)E-1} & $\mathbf{63(2)}$ & \num{210(7)} & \num{120(4)}\\
    (c) & \num{1.9E-6} & \num{4.95(6)E-3} & $\mathbf{268(10)}$ & \num{893(33)} & \num{510(19)}\\
    (d) & \num{1.5E-6} & \num{4.149(6)E-3} & $\mathbf{380(17)}$ & \num{1267(58)} & 724(33)\\
    \bottomrule[1.5pt]
\end{tabular}
\caption{Results of the ringdown measurements shown in Fig.~\ref{fig:ringdown}.}
\label{tab:results}
\end{table*}
 We repeated similar measurements for even lower pressures [see Figs.~\ref{fig:ringdown}(c) and (d)], down to $\SI{1.5e-6}{\milli\bar}$, where we reached the
%At low pressures approaching~$10^{-6}~{\rm mbar}$,  both the damping time $\gamma_3^{-1}$ and coupling rate $g(0)$ become even larger. Consequently, in the experiments illustrated in Figs.~\ref{fig:ringdown}(c) and (d), a waiting period of several hundred seconds is necessary for the spinning motion of the dumbbell to slow down sufficiently and make the precession mode frequency detectable.
% The damping and coupling rate parameters obtained from fits to the data shown in Figs.~\ref{fig:ringdown}(a)-(d) are presented in Table~\ref{tab:results}.
exceptionally high coupling rate of
$g(0)/\Omega_0 = 724(33)$, corresponding to a spinning rate of  ${\dot{\psi}(0)=2\pi \times \SI{1.20(6)}{GHz}}$. Notably, these low pressure ringdown measurements require a slowing down period of several hundred seconds for the precession mode frequency to become detectable [see Figs.~\ref{fig:ringdown}(c) and (d)]. This delay is a consequence of the  damping rate $\gamma_3$ decreasing to values below $10$~mHz and the spinning rate $\dot{\psi}(0)$ entering the GHz regime. 
%Table I lists our measured damping rates $\gamma_3$ for four selected pressures p_{gas}.
%The derived values of g, \dot{\phi} and g/\Omega_0 at time t=0 are listed as well.
%, achieved in the measurement carried out at~\SI{1.5e-6}{\milli\bar}.

\begin{figure}[t]
    \centering
    \includegraphics[width=0.4\textwidth]{"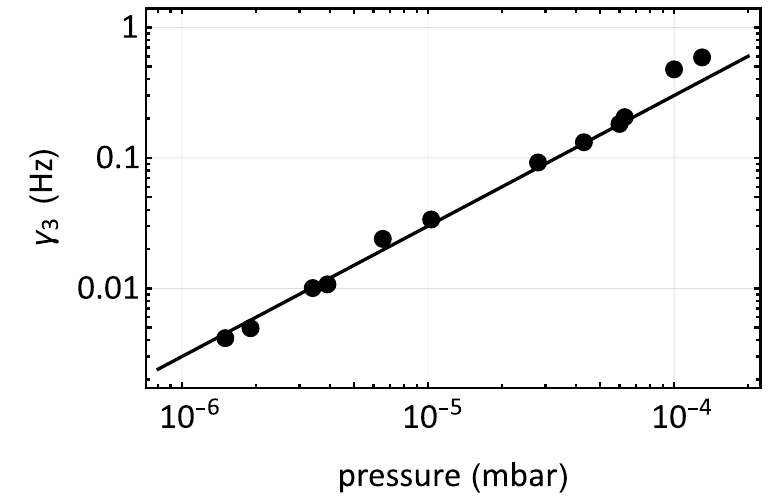"}
    \caption{Damping rate $\gamma_3$ as a function of pressure extracted from ringdown measurements. The solid line is a linear fit (forced through the origin).}
    \label{fig:damping}
\end{figure}

%Furthermore, combining the ringdown measurements at different pressures allows us to investigate the dependence of the damping rate $\gamma_3$ on pressure, shown in Fig.~\ref{fig:damping}. We conclude the damping rate $\gamma_3$ to be proportional to pressure, indicated by the linear fit shown as a solid line. Theory predicts~\cite{MartinetzPRE2018} that the damping rate for a symmetric rotor rotating around its axis of symmetry is proportional to a temperature-dependent accommodation coefficient as well as the gas density. Our measurements indicate that the accomodation coefficient remains constant in the pressure range between $10^{-6}$ and $10^{-3}$~mbar, implying that the internal temperature of the dumbbell throughout.

In Fig.~\ref{fig:damping} we plot our measurements of the damping rate $\gamma_3$ for different pressures together with a linear fit. According to theory, the damping of a symmetric rotor spinning along its long axis is proportional to a temperature-dependent accommodation coefficient as well as the gas density~\cite{MartinetzPRE2018}. Our measurements indicate that the accommodation coefficient remains constant in the pressure range between $10^{-6}$ and $10^{-3}$~mbar, implying that the internal temperature of the dumbbell remains constant in this pressure range.

\section*{Conclusions.} 
We have demonstrated, for the first time, controlled spinning of an optically levitated nanodumbbell around its long axis. %as a novel tool for rotational levitodynamics. 
%This was achieved by trapping the dumbbell with light in a 3D polarization state. At low pressures, we experimentally demonstrated controlled spinning of the nanodumbbell with a 
We achieve rotational rates exceeding $2 \pi \times 1\; \rm{GHz}$. The long axis spinning couples the two libration modes and the coupling rate is shown to be 724 times larger than the natural libration frequency.
%This spinning introduces a coupling rate between the two libration modes 724 times larger than the natural libration frequency.
This demonstration is an important step towards the development of levitated gyroscopes. 
%With this demonstrated technological advancement, we have taken an essential step towards developing levitated nanoparticles as gyroscopes. 
%This application requires an efficient detection of the spinning axis orientation, which is not feasible when the particle spins around its short axis~\cite{TebbenjohannsPRA2022}.
%Fast-spinning levitated nanoparticles offer significant advantages for inertial sensing of rotations.
%These objects, isolated from the surrounding environment, can be controlled down to their fundamental quantum and thermodynamic limits~\cite{vanderLaanPRA2020,vanderLaanPRL2021}.In contrast, conventional mechanical gyroscope technologies often face limitations due to technical issues.%
The GHz spinning rate demonstrated in this work translates into a thermally limited gyroscope sensitivity (angular random walk) of $\Omega_{\rm ARW}=\sqrt{4 k_B T \gamma}/(\sqrt{I_3} \dot{\psi}_0)\approx \SI{4e-6}{\radian \per \s \per \sqrt{\Hz}}$ for current experimental parameters~\cite{Poletkin2021}. This is only two orders of magnitude larger than the self-noise of the best high-end commercially available gyroscopes~\cite{blueseis}. 
Moreover, our experiment can be further optimized, e.g., by using larger dumbbells ($\Omega_{\rm ARW}\propto r^{-5/2}$, where $r$ is the nanoparticle radius) or by lowering the pressure below $10^{-6}\;\rm{mbar}$. 

%Even though our system is currently fully described classically, it is clear that r
Recent advances~\cite{ReimannPRL2018, AhnPRL2018, AhnNatNano2020, vanderLaanPRA2020, BangPRR2020, vanderLaanPRL2021, SchaeferPRL2021, BlakemorePRL2022, Xu2017PRA, RashidPRL2018, ju2023nearfield, SebersonPRA2019, BangPRR2020, vanderLaanPRL2021, kamba2023nanoscale, Pontin2023} have brought rotational levitodynamics close to the quantum regime~\cite{GonzalezBallesteroScience2021, SticklerNature2021}.
The deep-strong coupling regime shown in our work is an important step for realizing 
%induced by spinning motion may facilitate the observation of 
quantum correlations between ground-state cooled libration modes, as ground states of such strongly coupled systems are intrinsically entangled and contain virtual excitations~\cite{FriskKockum2019}. Our work further expands the potential that rotational degrees of freedom offer for the creation and measurement of macroscopic quantum states~\cite{SticklerNJP2018, Ma2020PRL, SticklerNature2021}.
%shows that rotational degrees of freedom may enable, in the near future, the creation and measurement of macroscopic quantum states~\cite{SticklerNJP2018, Ma2020PRL, SticklerNature2021}.
%This further expands the potential that rotational degrees of freedom offer for observing quantum effects on a macroscale~\cite{SticklerNJP2018, Ma2020PRL, SticklerNature2021}.    

%In particular, control over long axis rotation spin-rotational coupling effects, where magnetization can be converted into mechanical rotation (Barnett~\cite{Barnett}, Einstein de Haas~\cite{EinsteindeHaas})

%The ultrastrong coupling regime is reached as $g$ becomes comparable to the frequencies $\Omega_0$ of the uncoupled sub-components.

%Also, this work explores three-dimensional light polarization structure in levitodynamics. 

The authors would like to thank C. Gonzalez-Ballestero, J. Harris and the members of the Photonics Laboratory for fruitful discussions. This research was supported by the European Union’s Horizon 2020 research and innovation programme under grant agreement No.~[863132] (IQLev), as well as the ETH Grant No.~ETH-47 20-2. A. Norrman acknowledges support from the Research Council of Finland (Grant Nos. 354918 and 320166). F. van der Laan was supported by the Netherlands Organisation for Scientific Research (NWO).

\bibliography{spinbib}

\newpage
 
\section*{Methods}
\subsection*{Particle description}

In this work we focus on nanodumbbell particles, which are cylindrically symmetric objects comprised of two spheres attached to each other. We assume that the moment of inertia tensor as well as the real and imaginary parts of the polarizability tensor can be simultaneously diagonalized in the principal axes reference frame of the particle. We refer to this frame of reference as the "particle frame" and represent the principal axes by the unit vectors $(\mathbf{e}_1, \mathbf{e}_2, \mathbf{e}_3)$. We further let $\boldsymbol{\alpha'}=\text{diag} (\alpha_1',\alpha_1',\alpha_3')$ and $\boldsymbol{\alpha}''=\text{diag} (\alpha_1'',\alpha_1'',\alpha_3'')$ denote the real and imaginary parts of the static polarizability tensor of the object in the intrinsic body frame, respectively, assuming $\alpha_1' <\alpha_3'$. The moment of inertia of the particle is described by the tensor $\boldsymbol{I}= \text{diag} (I_1,I_1,I_3)$, where we assume $I_1 >I_3$. We refer to $\mathbf{e}_3$ (the principal axis with the largest polarizability) as the "long axis" of the object.

In order to describe the orientation of the particle frame with respect to the laboratory frame ($x,y,z$), we use the intrinsic $x$-convention of Euler angles denoted as $\phi$, $\theta$ and $\psi$ (see \cite{WolframEuler} and \S 35 in~\cite{Landau1976Mechanics}). The Euler angles $\phi$ and $\theta$ describe the orientation of the long axis of the rotor. In turn, the Euler angle $\psi$ describes rotation of the nanodumbbell around its long axis, i.e., $\dot{\psi}$ is the spinning speed. The angle measured in the experiment is $\phi$, which corresponds to the orientation of the long axis in the $xy$ plane with respect to the $y$ axis.

In order to transform a vector from the laboratory to the particle frame we first rotate it by the angle $\phi$ around $z$, then by $\theta$ around $\mathbf{e}_1$ and finally by $\psi$ around $\mathbf{e}_3$. The transformation matrix $\boldsymbol{R}$ corresponding to these three rotation operations is given in~\cite{WolframEuler}.

\subsection*{Conservative torques}
In this section we calculate the potential arising from the real part $\boldsymbol{\alpha}'$ of the polarizability tensor of the particle. The libration dynamics is dictated by the direction of the electric dipole moment induced by the optical field, whose orientation depends on the particle orientation and in general is not parallel to the electric field.

We describe the electric field of the linearly polarized trapping beam at the tweezers focal point as $\mathbf{E}_{\rm t}=(E_{\rm t} e^{i\omega_{\rm t}t}, 0, 0)^T$. In turn, we can write the electric field of the spinning beam (also at the tweezers focal point) as $\mathbf{E}_{\rm s}=(0, E_{\rm s} e^{i\omega_{\rm s}t}, i E_{\rm s} e^{i\omega_{\rm s}t})^\mathrm{T}$. 
The dipole moment induced on the trapped anisotropic particle, expressed in the laboratory frame, is given by
\begin{equation}
    \mathbf{p}=\boldsymbol{R^{-1} \alpha' R} \mathbf{E}, 
\end{equation}
where $\mathbf{E}=\mathbf{E}_{\rm t}+\mathbf{E}_{\rm s}$ is also expressed in the laboratory reference frame.
Since in general $\mathbf{p}$ and $\mathbf{}{E}$ are not parallel, the potential energy $ U$ associated with the orientation of the particle (after averaging over optical frequencies) is 
\begin{equation}
    U=-\frac{1}{4}\text{Re} \left(\mathbf{p}\cdot \mathbf{E}^*\right).
\end{equation}
Due to the fact that the trapping and spinning fields oscillate at different optical frequencies $\omega_{\rm t}$ and $\omega_{\rm s}$, corresponding to their respective wavelengths $\lambda _{\rm t}=1550$~nm and $\lambda _{\rm s}=1064$~nm, we can average out the cross terms oscillating at $\omega_{\rm s}-\omega_{\rm t}$ and calculate the potential components arising from both components independently. The potential arising from the trapping beam then reads
\begin{equation}
    U_{\rm t}=-\frac{1}{4} E_{\rm t}^2 \left(\alpha_1' + \Delta \alpha'  \sin \phi^2 \sin \theta^2\right),
\end{equation}
where $\Delta \alpha'=\alpha_3'-\alpha_1'$. The term $U_{\rm t}$ aligns the long axis of the particle with the trapping beam electric field, which points along $x$ axis. The potential arising from the spinning beam is
\begin{equation}
    U_{\rm s}=-\frac{1}{4} E_{\rm s}^2 \left(\alpha_1 '+ \alpha_3'-\Delta \alpha' \sin \phi^2 \sin \theta^2\right) .
\end{equation}
In turn, the potential term $U_{\rm s}$ tries to align the long axis of the particle to the $yz$ plane, thus counteracting $U_{\rm t}$. However, since the trapping field is much stronger than the spinning beam, $|E_{\rm t}|^{2}\gg |E_{\rm s}|^{2}$, we can approximate the total potential as $U\approx U_{\rm t}$. 

The long axis of the dumbbell oscillates in the potential minimum, occurring for $\phi=\theta=\pi/2$, which gives rise to two libration modes $\vartheta=\theta-\pi/2$ and $\varphi=\phi-\pi/2$. Expanding the potential around minimum and removing orientation-independent terms leads to
 \begin{equation}
    U \approx \frac{1}{2} I_1 \Omega_0^2 \; (\varphi^2 + \vartheta^2),
\end{equation}
where $\Omega_0=\sqrt{\Delta \alpha E_{\rm t}^2/2I_1}$ denotes the libration frequency. 

Finally, let us describe the libration dynamics due to the potential $U$ in terms of the restoring torque $\boldsymbol{\kappa}$,
whose components in the particle frame read
%
%\begin{subequations}
%\begin{align}
%   K_1 &= -\cos\psi \; \partial_\theta U-\sin\psi \frac{\partial_\phi U-\partial_\psi U \cos\theta}{\sin\theta}\\
%   K_2 &= \sin\psi \; \partial_\theta U-\cos\psi \frac{\partial_\phi U- \partial_\psi U\cos\theta}{\sin\theta}\\
%    K_3 &=-\partial_\psi U
%    \label{eq:torqueC1}
%\end{align}
%\end{subequations}
\begin{subequations}
\begin{align}
    \kappa_1 &= - I_1 \Omega_0^2 (\vartheta\cos \psi  +\varphi\sin \psi ),\\
    \kappa_2 &=- I_1 \Omega_0^2 (-\vartheta\sin \psi +\varphi\cos\psi),\\
    \kappa_3 & =0.
    \label{eq:torqueC2}
\end{align}
\end{subequations}
Note that the restoring torque components in the particle frame of reference depend on angle $\psi$, describing the rotation of the particle around its long axis.

\subsection*{Non-conservative torque}

The non-conservative dynamics in our experiment arises from the circular polarization of the spinning beam, which provides constant torque spinning the particle around its long axis. This torque can arise from optical absorption or scattering~\cite{BellandoPRL22} and depends on the $x$ component of the optical spin $S_x=\mathrm{Im}\left<\mathbf{E}^*\times\mathbf{E}\right>_x$ of the focused electric field $\mathbf{E}$ at the location of the dumbbell. The torque from absorption relies on the imaginary part of the dumbbell polarizability and is given by $\tau_{\rm abs}=\frac{1}{2}\alpha_3''S_x$.

Optical torque transfer by scattering requires some optical anisotropy that breaks the dumbbell's cylindrical symmetry. The scattering torque is given by $\tau_\mathrm{sc}=\frac{1}{2}(\delta\alpha')^2 g S_x$~\cite{BellandoPRL22}, where $\delta\alpha'$ is a measure for the optical anisotropy and $g=\omega^3/(6\pi\epsilon_0 c^3)$. This anisotropy could arise from a deviation from the perfectly spherical shape of the two particles constituting the dumbbell~\cite{kamba2023nanoscale}. 

Regardless of the mechanism of the angular momentum transfer from the light to the particle, we describe the effect of the circularly polarized spinning beam propagating along $x$ as a constant torque $\boldsymbol{\tau}= (\tau,0,0)$ in the lab frame. We expect the torque magnitude to be proportional to the spinning beam power, in other words $\tau\propto |E_{\rm s}|^2$. The components of $\mathbf{\tau}$ in the particle frame read:
%\begin{subequations}
%\begin{align}
%    \kappa_1 &= (\cos\phi \cos\psi - \cos\theta \sin\phi\sin\psi) k \nonumber\\ &\approx (-\varphi \cos \psi+\vartheta \sin\psi)k \\
%    \kappa_2 &=(-\cos\phi\sin\psi -\cos\theta\sin\phi\cos\psi)k \nonumber\\& \approx (\vartheta\cos \psi+\varphi \sin\psi)k\\
%    \kappa_3 & =( \sin\theta \sin\phi)k \approx k \;,
%    \label{eq:torqueNC}
%\end{align}
%\end{subequations}
\begin{subequations}
\begin{align}
    \tau_1 &=  (-\varphi \cos \psi+\vartheta \sin\psi) \tau ,\label{eq:torquetau1}\\
    \tau_2 &= (\vartheta\cos \psi+\varphi \sin\psi) \tau ,\label{eq:torquetau2}\\
    \tau_3 & = \tau \;,
    \label{eq:torqueNC}
\end{align}
\end{subequations}
where we have expanded to the first order around the particle orientation in the potential minimum (set by the trapping beam). 

Therefore, the total torque ${\mathbf{K}}$ experience by our particle, which includes both the conservative torque arising from the potential $U$ and non-conservative spinning torque ${\boldsymbol{\tau}}$ amounts to
\begin{subequations}
\begin{align}
    K_1 &=\tau_1+\kappa_1 \approx  - I_1 \Omega_0^2 (\vartheta\cos \psi  +\varphi\sin \psi ), \label{eq:torqueFull1}\\
    K_2 &=\tau_2+\kappa_2 \approx - I_1 \Omega_0^2 (-\vartheta\sin \psi +\varphi\cos\psi),  \label{eq:torqueFull2}\\
    K_3 & =\tau_3+\kappa_3 \approx \tau.
    \label{eq:torqueFull3}
\end{align}
\end{subequations}
when expressed in the particle frame of reference. We have neglected the terms $\tau_1$ and $\tau_2$ [see Eqs.~\eqref{eq:torquetau1} and ~\eqref{eq:torquetau2}], which introduce a small coupling between libration modes $\varphi$ and $\vartheta$. This coupling has a negligible effect on the libration dynamics (dominated by the strong coupling introduced by the spinning motion), as $|E_{\rm s}|^2 \ll |E_{\rm t}|^2$ implies that $\tau\ll I_1 \Omega_0^2$. 

\subsection*{Equations of motion}

This section contains the analysis of the Euler equations of motion in presence of the torque derived in the previous section. We begin by expressing the angular velocity ${\bf \omega}$ in the particle frame and expanding it to the first order in libration angles $\varphi$ and $\vartheta$: 
\begin{subequations}
\begin{align}
    \omega_1 &\approx\dot{\varphi}\sin\psi+\dot{\vartheta}\cos\psi \label{eq:omegas1},\\
    \omega_2 &\approx \dot{\varphi} \cos\psi -\dot{\vartheta}\sin\psi \label{eq:omegas2},\\
    \omega_3 &\approx -\dot{\varphi}\vartheta+ \dot{\psi}.
    \label{eq:omegas3}
\end{align}
\end{subequations}
In the remainder of this work we are interested in the regime of large $\dot\psi$ and small-amplitude libration, therefore we further approximate the third component of the angular velocity as $\omega_3\approx \dot\psi$. 

\begin{figure}[t]
    \centering
    \includegraphics[width=0.4\textwidth]{"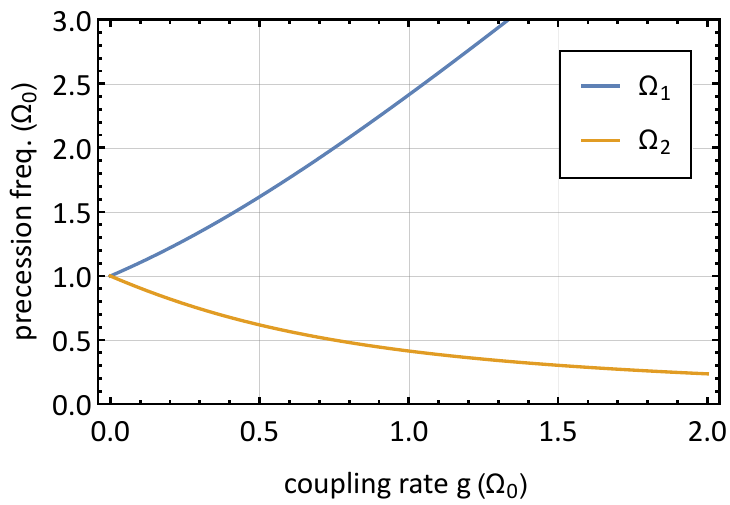"}
    \caption{Precession modes $\Omega_1$ and $\Omega_2$ calculated from Eq.~\eqref{eq:precfrs} in the units of natural libration frequency $\Omega_0$.}
    \label{fig:prectheosup}
\end{figure}

Particle-frame Euler equations of motion for our cylindrically symmetrical dumbbell read: 
\begin{subequations}
\begin{align}
   K_1&=I_1 \dot{\omega_1} -\Delta I \omega_2 \omega_3, \label{eq:motion1}\\
   K_2&=I_1 \dot{\omega_2} +\Delta I \omega_1 \omega_3, \label{eq:motion2}\\
   K_3&=I_3 \dot{\omega_3},
    \label{eq:motion3}
\end{align}
\end{subequations}
where $\Delta I=I_1-I_3$. Using the expressions for torque (see Eqs.~\eqref{eq:torqueFull1}--\eqref{eq:torqueFull3}) and angular velocity (see Eqs.~\eqref{eq:omegas1}--\eqref{eq:omegas3}) components, we can rewrite Eqs.~\eqref{eq:motion1}--\eqref{eq:motion3} as: 
\begin{subequations}
\begin{align}
   I_1&  \ddot{\vartheta}+ I_3 \dot{\varphi}\dot{\psi}=  - I_1 \Omega_0^2 \vartheta \;,\label{eq:motion21}\\
   I_1&  \ddot{\varphi} -I_3 \dot{\vartheta} \dot{\psi}= - I_1 \Omega_0^2 \varphi\;,\label{eq:motion22}\\
   I_3&  \ddot{\psi} = \tau.
    \label{eq:motion23}
\end{align}
\end{subequations}
Note that Eqs.~\eqref{eq:motion21} and~\eqref{eq:motion22} describe two coupled harmonic oscillators $\vartheta$ and $\varphi$, whereas according the Eq.~\eqref{eq:motion23}, the velocity of the spinning around the long axis $\dot{\psi}$ increases to infinity. In order to avoid this problem, we have to expand the model to include friction in Eq.~\eqref{eq:motion23}. In presence of friction (described by coefficient $\gamma_3$), the spinning speed $\dot{\psi}$ will increase until it reaches a stationary value $\dot{\psi}_{0}=\tau/(I_3\gamma_3$), for which driving torque $\tau$ is balanced by the friction~\cite{ReimannPRL2018, AhnPRL2018}.

\subsection*{Coupled libration modes}

Let us now focus on the coupled libration modes described by Eqs.~\eqref{eq:motion21} and~\eqref{eq:motion22} and rewrite them explicitly introducing the coupling rate $g$, 
\begin{subequations}
\begin{align}
 \ddot{\vartheta} +2g \dot{\varphi}=  -  \Omega_0^2 \vartheta \;,\\
  \ddot{\varphi} -2g \dot{\vartheta} = - \Omega_0^2 \varphi\;,
    \label{eq:coupled}
\end{align}
\end{subequations}
where $g=(I_3/2I_1) \dot{\psi}_{0}$. In agreement with \cite{SebersonPRA2019}, the solution can be written as 
%$\vartheta = {\mathcal Re} (\Theta) $ and $\varphi={\mathcal Re} (\Phi)$
%\begin{subequations}
%\begin{align}
%\Theta&= {\left( A e^{i \Omega_1 t}+B e^{i\Omega_2 t}\right)}\;,\\
%\Phi&={\left(-i A e^{i \Omega_1 t}+ i B e^{i\Omega_2 t}\right)}\,
%\end{align}
%\end{subequations}
\begin{subequations}
\begin{align}
\vartheta&=  \alpha_1 \cos { (\Omega_1 t+\delta_1)}+ \alpha_2 \cos { (\Omega_2 t+\delta_2)}\;,\\
\varphi&=  \alpha_1 \sin { (\Omega_1 t+\delta_1)}- \alpha_2 \sin { (\Omega_2 t+\delta_2)}
\label{eq:sol2}
\end{align}
\end{subequations} 
where $\alpha_{1/2}$ and $\delta_{1/2}$ depend on initial angular displacements and velocities and the eigenfrequencies read 
\begin{equation}
\Omega_{1/2}=\sqrt{\Omega_0^2+g^2}\pm g .
\label{eq:Supprecfr}
\end{equation} 
Note that the fast precession (nutation) mode with frequency $\Omega_1$ corresponds to a circular motion of the long axis in the counterclockwise direction (for $\boldsymbol{\tau}$ pointing along the positive $x$ direction), and the slow precession mode corresponds to a clockwise long axis motion.

The dependence of $\Omega_{1/2}$ on the coupling rate $g$ is shown in Fig.~\ref{fig:prectheosup}. For a fast spinning dumbbell ($g \gg\Omega_0$) the frequencies can be approximated as:
\begin{subequations}
\begin{align}
\Omega_{1}&= 2g \;,\\
\Omega_2&=\frac{\Omega_0^2}{2g} \label{eq:slow mode}.
\end{align}
\end{subequations}
Finally, let us turn our attention to the amplitudes of precession and nutation. For a rapidly spinning top, the ratio between the amplitude of nutation $\alpha_1$ and the amplitude of precession $\alpha_2$ decreases proportionally to the spinning speed $\dot{\psi}$ squared~\cite{Goldstein}. Therefore we expect nutation will be negligible for a rapidly spinning dumbbell and the dynamics will be dominated by slow precession with frequency $\Omega_2$. The amplitude of the slow precession $\alpha_2$ is equivalent to the tilt between the long axis of the dumbbell and $x$ axis. This angle will stay almost constant as the dumbbell precesses, except for fast, small-amplitude oscillations caused by nutation. The effective potential governing the motion of the tilt angle $\alpha_2$ can be written as $U_{\rm ef}=\frac{1}{2} I_1 (g^2+\Omega_0^2)\alpha_2^2$~\cite{Landau1976Mechanics}. Applying equipartition theorem to this degree of freedom allows us to predict the average precession amplitude (tilt angle) of the thermally driven spinning top, which yields:
\begin{equation}
\langle \alpha_2^2\rangle=\frac{k_B T}{I_1 (\Omega_0^2+g^2) }
\end{equation}
Comparing the above result with Eq.~\ref{eq:sol2} indicates that the oscillation amplitude of the libration angle $\varphi$ measured in our experiment will behave similarly (if the angular degrees of freedom except $\psi$ are in thermal equilibrium with the surrounding gas). To summarize, we expect that both the amplitude of the nutation and precession will diminish as we increase the spinning speed, with nutation signal diminishing much faster. In practice, the amplitude of precession motion may be affected by misalignment between the spinning beam and the tweezers' polarization and external rotation of the experimental apparatus (e.g. due to the floating optical table). 

\subsection*{The spinning torque}

We estimate that the magnitude of the spin vector $\mathrm{Im}\left<\mathbf{E}_s^*\times\mathbf{E}_s\right>$ \cite{GilPRA2023} carried by the spinning beam can reach up to $5 \times10^{12} \;{\rm V^2/m^2}$. For comparison, the electric field squared corresponding to the trapping beam at the focal spot is 200 times larger. Additionally, it is interesting to compare the spin angular momentum carried by the spinning field to the transverse spin generated by focusing the linearly polarized trapping beam near the trapping region~\cite{novotny_hecht_2012,ZielinskaPRL2023}-- in our experiment the angular momentum of the spinning beam is an order of magnitude smaller.

The damping rates $\gamma_3$ and the spinning rates extracted from the ringdown measurements (see main text) allow us to estimate the torque exerted on the dumbbell by a circularly polarized beam, arriving at $\tau=\SI{5e-25}{\N\m}$ for \SI{100}{\mW} of the optical power. We are unable to determine whether the torque arises from optical absorption, scattering (which requires breaking the cylindrical symmetry of the dumbbell) or other processes~\cite{Toftul2023}. Attributing this torque to absorption requires the dumbbell's absorption coefficient to be two orders of magnitude larger than expected for bulk silica at $1064\;\rm{nm}$~\cite{KitamuraApplOpt07}. On the other hand, attributing the torque to the scattering process~\cite{BellandoPRL22}, requires the polarizability of the dumbbell along the two short axes to differ by approximately $3\%$. Both increased absorption and imperfect spherical shape of the silica nanoparticles have been reported by other researchers~\cite{MonteiroPRA2018,kamba2023nanoscale}.

\subsection*{Particle deformation due to centrifugal forces}

The calculated tensile stress generated by centrifugal force in silica nanodumbbells in our experiments reaches $0.5~\rm{GPa}$~\cite{Schuck2018,ReimannPRL2018}. The high spinning rates generate a tensile stress of $0.5~\rm{GPa}$~\cite{Schuck2018,ReimannPRL2018} and approach the regime of deforming the particle shape~\cite{Hummer2020}. However, no consistent changes in characteristic frequencies and damping rates, temporary or permanent, were observed. We have recorded a single case of deformation (out of $\sim15$ dumbbells investigated in total). After spinning with approx. $200\; \rm{MHz}$, the particle's $x$-to-$y$ gas damping ratio changed from 1.14 to 1.07, and the natural libration frequency decreased by $35\%$. The data obtained from this particle is not used in the manuscript.

\end{document}